\documentclass[prl,twocolumn,groupedaddress,amssymb,amsmath,amsfont,showpacs,checklength]{revtex4}
\usepackage{bm}
\usepackage{amsmath}
\usepackage{amssymb}
\usepackage{amsthm}
\usepackage{amsfonts}
\usepackage{graphicx}
\usepackage{color}
\usepackage[colorlinks=true,urlcolor=blue,linkcolor=green,citecolor=red]{hyperref}

\begin{document}


\title{Correlative Capacity of Composite Quantum States}


\author{M. Hossein Partovi}
\email[Electronic address:\,\,]{hpartovi@csus.edu}
\affiliation{Department of Physics and Astronomy, California State
University, Sacramento, California 95819-6041}


\date{\today}

\begin{abstract}
We characterize the optimal correlative capacity of entangled, separable, and classically correlated states. Introducing the notions of the \textit{infimum} and \textit{supremum} within majorization theory, we construct the least disordered separable state compatible with a set of marginals.  The maximum separable correlation information supportable by the marginals of a multi-qubit pure state is shown to be an LOCC monotone.  The least disordered composite of a pair of qubits is found for the above classes, with classically correlated states defined as diagonal in the product of marginal bases.

\end{abstract}


\maketitle



The development of statistical ideas in the nineteenth century underscored the importance of correlations for composite systems \cite{hp}.  A fundamental tenet of the statistical description was the notion that the uncertainty in the knowledge of the whole cannot be smaller than that of any of its parts, and that a maximal knowledge of the whole guarantees the same for its parts.  The discovery of quantum mechanics revealed the astonishing possibility that parts of a maximally known microscopic system can be partially or even totally unknown, or rather \textit{unknowable}, thus signaling an extraordinary level of correlations.  Such perplexing correlative features were noted early and named ``entanglement'' \cite{Ein/Sch}, and have since been established as essential to the structure of quantum mechanics \cite{Wer/Wol}.  They arise from an interplay of superposition and correlation in a composite quantum system \cite{Rec}.  The recognition that superposed correlations are experimentally distinguishable from classically feasible ones came with the celebrated work of Bell who exploited a   fundamental incompatibility between them \cite{Bell}.  The remarkable progress of quantum information science during the past three decades may be viewed as a systematic application of the correlative power of microscopic systems to achieving classically impossible or inefficient tasks \cite{Ben,NC}.

An important step in the characterization of correlations was the identification of entanglement with inseparability, i.e., the impossibility of representing a composite state as a mixture of pure product states \cite{Wer}.  Indeed so defined, entanglement is the crucial ingredient for realizing the exotic phenomena of quantum information processing \cite{Ben,NC}.  However, nonclassical behavior is by no means limited to entangled states, a fact that was recognized early in relation to quantum nonlocality \cite{Ben/eta} and has been further emphasized in recent years \cite{Lit}, especially in regard to quantum computing.  Thus separable quantum states, although often casually characterized as ``classically correlated,'' may possess properties that are classically impossible.  Furthermore, while entanglement is subject to saturation in a multipartite system (``monogamy''), separable correlations are not so limited.  In addition, separable correlations are expected to be less fragile against environment-induced decoherence, the dreaded \textit{b\^{e}te noire} of quantum information processing.  Finally, separable correlations can be generated by local operations and classical communication (LOCC), which is their defining characteristic \cite{Wer}.   Thus there are compelling reasons for viewing separable correlations as a potential resource, and studying them as a means of gaining a better understanding of the structure of correlations in quantum mechanics.

In this Letter we address the problem of characterizing the correlative capacity of separable quantum states, including classically correlated ones.    Introducing an extension of majorization relations, we develop a complete solution to the quantum marginal problem of determining the least disordered separable state compatible with a given set of marginals.  This solution has a separable form in its principal ensemble representation.  We interpret the total correlation content of this state as the maximum possible for any state that is marginally isospectral with it, and show that it is an LOCC monotone as well as an entanglement measure for any pure, multi-qubit state.  These general results are illustrated by finding and ranking the maximal correlation information for entangled, separable, and classically correlated composites of a pair of qubits.  We define classically correlated states as those that are diagonal in the product of their marginal bases, and find this to coincide with the symmetrized version of the definition based on quantum discord \cite{cfs}.

Let $\mathfrak{M}$ be a set of local states ${\rho}^{a} \in {\mathcal{H}}^{a},\, a=1,2,\ldots, N$.  Our objective is to construct the least disordered separable state compatible with $\mathfrak{M}$.  We will accomplish this by introducing the notion of the \textit{infimum} of a set of density matrices within majorization theory \cite{maj}.   The majorization relation is a partial order on real vectors that serves to compare the degree of disorder among those that are comparable.  Because the vectors in our applications will be the spectra of density matrices, we will restrict our attention to vectors with non-negative components arranged in a descending order and summing to unity.  Given a pair of such vectors ${\lambda}^{1}$ and ${\lambda}^{2}$, ${\lambda}^{1}$ is said to be majorized by ${\lambda}^{2}$ and written as ${\lambda}^{1}\prec{\lambda}^{2}$ if ${\sum}_{i}^{j} {\lambda}^{1}_{i} \leq {\sum}_{i}^{j} {\lambda}^{2}_{i} $ for $j=1,2, \ldots d$, where $d$ is the larger of the two dimensions and trailing zeros are added where needed.  Likewise, given a pair of density matrices ${\rho}^{1}$ and ${\rho}^{2}$, we say ${\rho}^{1}$ is majorized by ${\rho}^{2}$ and write ${\rho}^{1}\prec{\rho}^{2}$ if $\lambda({\rho}^{1})\prec \lambda({\rho}^{2})$.  Here $\lambda({\rho})$ denotes the spectrum of $\rho$.

The infimum of a set of $N$ vectors is now defined as the vector that is majorized by every element of the set and in turn majorizes any vector with that property. To implement this definition, consider a vector $\mu$ such that ${\mu}_{0}=0$ and ${\mu}_{j}=\min({\sum}_{i=1}^{j} {\lambda}^{1}_{i},{\sum}_{i=1}^{j} {\lambda}^{2}_{i},\ldots, {\sum}_{i=1}^{j} {\lambda}^{N}_{i}),\,\, j=1,2,\ldots,{d}_{max}$, where ${d}_{max}$ is the largest dimension occurring in the set.  The desired infimum is now given by
\begin{equation}
{\inf({\lambda}^{1},{\lambda}^{2}, \ldots, {\lambda}^{N})}_{i}={\mu}_{i}-{\mu}_{i-1},\,i=1,2,\ldots,{d}_{max}. \label{1}
\end{equation}
Similarly, a density matrix is said to be the infimum of a set of density matrices if its spectrum is the infimum of those in the set.  The supremum of a set of vectors is defined in a similar manner, namely, as the vector that majorizes every element of the set and is in turn majorized  by any vector with that property \cite{note}.  The constructive definition of the supremum is slightly more involved and is given elsewhere \cite{code}.

Note that the majorization relation as a comparator of disorder is in general stronger than entropy, i.e., ${\rho}^{1} \prec {\rho}^{2}$ implies $S({\rho}^{1}) \geq S({\rho}^{2})$ but not conversely, where $S(\cdot)$ is the von Neumann entropy.  Note also an important classification theorem formulated in majorization terms by Nielsen \cite{Nls1}:  Given a vector $\Lambda$ and a density matrix $\rho$, the ensemble representation $\rho={\sum}_{\alpha} {\Lambda}_{\alpha} |{\phi}_{\alpha} \rangle\langle{\phi}_{\alpha}|$ exists if and only if $\Lambda \prec \lambda(\rho)$.

We are now in a position to construct the least disordered separable state compatible with the set of marginals $\mathfrak{M}$ defined earlier.  Let $\Lambda$ be defined as
\begin{equation}
\Lambda=\inf[{\lambda}({\rho}^{1}), {\lambda}({\rho}^{2}), \ldots, {\lambda}({\rho}^{N})].  \label{2}
\end{equation}
Then, using the above theorem, we can represent the $a$th marginal state as ${\rho}^{a}={\sum}_{\alpha} {\Lambda}_{\alpha}|{\phi}^{a}_{\alpha} \rangle \langle{\phi}^{a}_{\alpha}|$, where the vectors $\{ |{\phi}^{a}_{\alpha} \rangle \}$ are normalized to unity and $1 \leq \alpha \leq d$, with $d$ representing the dimension of $\Lambda$.  Exploiting the fact that all local states are now represented by means of the same probability set $\Lambda$, we can  assemble the desired global state as
\begin{equation}
{\sigma}(\mathfrak{M})={\sum}_{\alpha} {\Lambda}_{\alpha} |{\phi}^{1}_{\alpha} \rangle\langle{\phi}^{1}_{\alpha}|\otimes|{\phi}^{2}_{\alpha} \rangle \langle{\phi}^{2}_{\alpha}|\otimes \ldots |{\phi}^{N}_{\alpha} \rangle\langle{\phi}^{N}_{\alpha}|.  \label{3}
\end{equation}
It is clear by inspection that the marginals of ${\sigma}(\mathfrak{M})$ are precisely those in the set $\mathfrak{M}$.

Next we will show that $\lambda[{\sigma}(\mathfrak{M})]=\Lambda$. To that end we recall another important result of majorization theory due to Nielsen and Kempe \cite{Nls2}: the spectrum of a separable density matrix is majorized by those of its marginals.  Applied to ${\sigma}(\mathfrak{M})$, this theorem implies that $\lambda[{\sigma}(\mathfrak{M})] \prec \lambda({\rho}^{a})$ for every $a$.  Since, by its construction in Eq.~(\ref{2}), $\Lambda$ majorizes every vector that is majorized by the marginal spectra of $\sigma$, it must majorize $\lambda[{\sigma}(\mathfrak{M})]$, i.e., $\lambda[{\sigma}(\mathfrak{M})] \prec \Lambda$.  The opposite of this  relation also holds as a consequence of the representation theorem cited above.  The antisymmetry property of the majorization relation then implies that $\lambda[{\sigma}(\mathfrak{M})]=\Lambda$.  Recall that the infimum property of $\Lambda$ guarantees that ${\sigma}(\mathfrak{M})$ majorizes any separable state isomarginal with it.  In other words, ${\sigma}(\mathfrak{M})$ is the least disordered separable state compatible with the marginal set $\mathfrak{M}$.

The equality $\lambda[{\sigma}(\mathfrak{M})]=\Lambda$ deduced above would follow if the representation of Eq.~(\ref{3}) were orthogonal.  We will next show that this is in fact the case.  The set of $d$ vectors $|{\phi}_{\alpha} \rangle=|{\phi}^{1}_{\alpha} \rangle \otimes |{\phi}^{2}_{\alpha} \rangle \otimes \ldots |{\phi}^{N}_{\alpha} \rangle$ in Eq.~(\ref{3}) spans the support of ${\sigma}(\mathfrak{M})$ which, by virtue of the equality $\lambda[{\sigma}(\mathfrak{M})]=\Lambda$, is a subspace of dimension $d$.  This guarantees the linear independence of the set $\{|{\phi}_{\alpha} \rangle \}$.  This property can in turn be used to show that the spectrum of ${\sigma}(\mathfrak{M})$ is the same as that of the Hermitian matrix ${G}_{\alpha \beta}={\Lambda}_{\alpha}^{1/2} \langle {\phi}_{\alpha} |{\phi}_{\beta} \rangle {\Lambda}_{\beta}^{1/2}$, i.e., $\lambda[{\sigma}(\mathfrak{M})]=\lambda(G)=\Lambda$.  Note that owing to the normalization condition $\langle {\phi}_{\alpha} |{\phi}_{\alpha} \rangle=1$, the diagonal elements of $G$ coincide with its spectrum.  The extremum property of Hermitian matrices can then be used to conclude that the off-diagonal elements of $G$ must be zero.  This establishes the orthogonality of the representation of Eq.~(\ref{3}).  In summary, we have

\textit{Theorem 1. The least disordered separable state compatible with a given set of marginals has a spectrum that is the infimum of its marginal spectra and a representation that is separable and orthogonal}.

An immediate corollary of this result is a sharpened version of the Nielsen-Kempe theorem \cite{Nls2}:

\textit{Corollary 1. The spectrum of a separable state is majorized by the infimum of its marginal spectra.}

Thus far we have used the majorization relation to characterize the least disordered state compatible with a set of marginals.  As a scalar measure of the correlation content of a state, we will use the following generalization of mutual information:
\begin{equation}
C({\rho})=S(\rho \|{\rho}^{1} \otimes {\rho}^{2} \otimes \ldots {\rho}^{N})= {\sum}_{\alpha=1}^{N}S({\rho}^{\alpha}) -S(\rho),  \label{4}
\end{equation}
where $\{ {\rho}^{\alpha} \}$ are the marginals of $\rho$ and $S(\cdot \| \cdot)$ is the quantum relative entropy function \cite{NC}.  Note that the order induced by $C(\cdot)$ is weaker than that implied by majorization, i.e., given a pair of isomarginal density matrices ${\rho}$ and ${\gamma}$, ${\rho} \prec {\gamma}$ implies $C({\rho}) \leq C({\gamma})$ (but not conversely).  Note also that the correlation information function $C(\cdot)$ is additive in the sense that, given any partitioning of a multipartite state $\rho$ into $M$ reduced states $\{ {\rho}_{i} \}$ (not necessarily the marginals), we have $C({\rho})={\sum}_{i} C({ \rho}_{i})+S(\rho \|{\rho}_{1} \otimes {\rho}_{2} \otimes \ldots {\rho}_{M})$.  Combining this measure with Theorem 1, we have

\textit{Corollary 2. The maximum correlation information that can be encoded by a separable state compatible with the set of marginal spectra ${ \{ {\lambda}^{a} \} }$ is
\begin{equation}
{\sum}_{a=1}^{N} H({\lambda}^{a})-H(\inf[{\lambda}^{1}, {\lambda}^{2}, \ldots, {\lambda}^{N}]), \label{5}
\end{equation}
where $H(\cdot)$ is the Shannon entropy function.}

At this juncture we must define the subset of separable states that we regard as ``classically correlated'' \cite{cfs}.   The physical basis of our definition is the requirement that the correlative structure of a classically correlated quantum state must be classically feasible.  More specifically, the spectrum of a classically correlated density matrix must correspond to a joint probability distribution whose (classical) marginals match the marginal spectra of the density matrix.  This requirement leads to

\textit{Definition 1. A quantum state is classically correlated iff its density matrix has a diagonal form in the product of its marginal bases.}

To see how this definition meets the above-stated requirement, consider the set of marginals $\mathfrak{M}$ together with the principal representations ${\rho}^{a}={\sum}_{i} {\lambda}^{a}_{i} {\Pi}^{a}_{i}$, where ${\lambda}^{a}_{i}$ and ${\Pi}^{a}_{i}$ are the marginal spectra and corresponding projection operators.  The above definition then requires that a classically correlated composite of $\mathfrak{M}$ must have the form ${\sum}{p}_{{i}_{1}{i}_{2} \ldots {i}_{N}} {\Pi}^{1}_{i_{1}}\otimes{\Pi}^{2}_{i_{2}}\otimes \ldots \otimes{\Pi}^{N}_{i_{N}}$, where a sum over all subscripts is implied.  Note that the classical marginals of the joint probability distribution ${p}_{{i}_{1}{i}_{2} \ldots {i}_{N}}$, which is the spectrum of the composite state, are precisely the marginal spectra of the composite as required.

The above definition is clearly more restrictive than the vanishing of the quantum discord (and its non-projective version) popular in the literature \cite{cfs}.  Remarkably, symmetrizing the vanishing condition of the discord with respect to all parties, which is a basic requirement classically, directly leads to our definition.  For example, for a bipartite composite $\rho$, the two vanishing conditions are ${\sum}_{i} {\Pi}^{1}_{i}\rho {\Pi}^{1}_{i}=\rho $ and ${\sum}_{i} {\Pi}^{2}_{i}\rho {\Pi}^{2}_{i}=\rho $, which jointly imply that $\rho$ has the desired structure with ${p}_{{i}_{1}{i}_{2}}={\textrm{tr}}({\Pi}^{1}_{{i}_{1}}\otimes{\Pi}^{2}_{i_{2}}\rho {\Pi}^{1}_{{i}_{1}}\otimes{\Pi}^{2}_{i_{2}})$.

As an illustration of the above ideas, we will construct and compare the least disordered composites of a pair of qubits ${\rho}^{a}$ and  ${\rho}^{b}$, with spectra ${\lambda}^{a}=({p}^{a},1-{p}^{a})$ and ${\lambda}^{b}=({p}^{b},1-{p}^{b})$, for classically correlated, separable, and entangled cases. The least disordered separable state ${\sigma}^{s}$ is determined by applying the construction of Eq.~(\ref{3}):
\begin{equation}
{\sigma}^{s}={p}^{b}|\theta 0 \rangle \langle \theta 0|+(1-{p}^{b})|11 \rangle \langle 11|,  \label{6}
\end{equation}
where $|\theta 0 \rangle =[\cos(\theta)|0 \rangle  +\sin (\theta)|1 \rangle ] \otimes |0 \rangle \ $, and   ${\cos(\theta)}^2={p}^{a}(1-{p}^{a})/{p}^{b}(1-{p}^{b})$.  Except for Fig. 1 below, ${p}^{a} \geq {p}^{b}$ is assumed without loss of generality.  Note that the representation of ${\sigma}^{s}$ in Eq.~(\ref{6}) is orthogonal as expected, with the corresponding spectrum given by ${\Lambda}^{s}=({p}^{b},1-{p}^{b},0,0)$.
\begin{figure}
\includegraphics[]{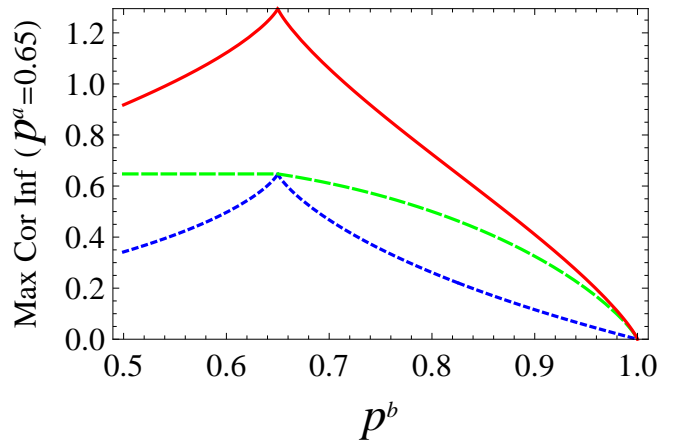}
\caption{Maximum correlation information for two-qubit, classically correlated (short dash, blue), separable (long dash, green), and entangled (solid, red).}
\label{fig}
\end{figure}
To find the least disordered entangled state ${\sigma}^{e}$ marginally isospectral with ${\sigma}^{s}$, we first consider its standard purification, which is a composite of two qubits and a 4-dimensional qudit.  As such, the spectra of the three marginals are subject to a number of conditions given in Theorem 3 of Ref. \cite{hig}, including a generalization of the so-called triangle inequalities $1-{p}^{b} \leq 1-{p}^{a} + 1-{\lambda}_{max}$, where ${\lambda}_{max}$ is the largest eigenvalue of ${\sigma}^{e}$.  We will see below that the bound on ${\lambda}_{max}$ implied by this inequality is in fact saturated, and that ${\Lambda}^{e}=(1+{p}^{b}-{p}^{a},{p}^{a}- {p}^{b},0,0)$, where ${\Lambda}^{e}$ is the spectrum of ${\sigma}^{e}$.

To construct the state ${\sigma}^{e}$, let the marginal states ${\rho}^{a}$ and ${\rho}^{b}$ be represented as ${p}^{a} |0 \rangle \langle 0|+(1-{p}^{a})|1 \rangle \langle 1|$ and ${p}^{b} |0 \rangle \langle 0|+(1-{p}^{b})|1 \rangle \langle 1|$, respectively.  Then, using the product basis $\{ |00 \rangle,|01 \rangle,|10 \rangle,|11 \rangle \}$, we find
\begin{equation}
{\sigma}^{e}=
\begin{pmatrix}
{p}^{b} & 0 & 0 & {[{p}^{b}(1-{p}^{a})]}^{1/2}\\
0 & {p}^{a}-{p}^{b} & 0 & 0\\
0 & 0 & 0 & 0\\
{[{p}^{b}(1-{p}^{a})]}^{1/2} & 0 & 0 & 1-{p}^{a}
\end{pmatrix}.
\label{7}
\end{equation}
A straightforward calculation shows that ${\sigma}^{e}$ given in Eq.~(\ref{7}) has ${\Lambda}^{e}$ as its spectrum and ${\rho}^{a}$ and ${\rho}^{b}$ as its marginals. Since ${\Lambda}^{e}$ is the least disordered possible, we conclude that ${\sigma}^{e}$ is indeed the desired optimal state.

Finally, the least disordered classically correlated state ${\sigma}^{c}$ is obtained by decohering ${\sigma}^{e}$ in Eq.~(\ref{7}).  Therefore, the (ordered) spectrum of ${\sigma}^{c}$ is given by ${\Lambda}^{c}=[{p}^{b},\max({p}^{a}-{p}^{b},1-{p}^{a}), \min({p}^{a}-{p}^{b},1-{p}^{a}),0]$.  One can verify that ${\Lambda}^{c}$ majorizes any joint probability distribution whose (classical) marginals are ${\lambda}^{a}$ and ${\lambda}^{b}$.

We summarize the above results as

\textit{Theorem 2.  Let ${\rho}^{c}$, ${\rho}^{s}$, and ${\rho}^{e}$ be, respectively, any classically correlated, separable, and entangled composite of the given qubits ${\rho}^{a}$ and ${\rho}^{b}$.  Then ${\rho}^{c}\prec {\sigma}^{c}$, ${\rho}^{s}\prec {\sigma}^{s}$, and ${\rho}^{e}\prec {\sigma}^{e}$, with $ \lambda ({\sigma}^{c})\prec \lambda({\sigma}^{s})\prec \lambda({\sigma}^{e})$, where $ {\sigma}^{c}$, ${\sigma}^{s}$, and ${\sigma}^{e}$ and their spectra are as given above}.

This theorem establishes a remarkable hierarchy of disorder among two-qubit states that are locally indistinguishable.  As an illustration, Fig. 1 shows a plot of maximum correlation information for the three cases, $C({\rho}^{c})$, $C({\rho}^{s})$, and  $C({\rho}^{e})$, versus ${p}^{b}$, with ${p}^{a}$ fixed at $0.65$ (an animated version is given in Ref. \cite{code}).  Note that in case the two qubits are isospectral (e.g., for ${p}^{b}=0.65$ in Fig. 1), ${\sigma}^{s}$ is classically correlated and ${\sigma}^{e}$ is pure.  The
latter generalizes to pure composites of $N$ isospectral qudits, which may be called \textit{feline} states (after Schr\"{o}dinger's cat \cite{Ein/Sch}).   These states admit a Schmidt decomposition and carry a correlation information of $N{S}_d$, where ${S}_d$ is the entropy of each qudit. Their decohered version is classically correlated and carries a correlation information of $(N-1){S}_d$, nearly the same as the pure state for $N \gg 1$, reflecting the saturation property of entanglement.

What is the LOCC behavior of the maximum separable correlation information encodable by the marginals of a pure state?  To answer this question, we  consider a pure state $\psi$ with the qubits $\{ {\rho}^{1}(\psi), {\rho}^{2}(\psi),\ldots,{\rho}^{N}(\psi)\}$ as its marginals.  Let $\sigma(\psi)$ be the least disordered separable state composed of these marginals, as in Eq.~(\ref{3}). What is the behavior of $C[\sigma(\psi)]$ under a sequence of local operations and classical communications that culminate in a set of pure states $\{ {\psi}_{i} \}$ with probabilities $\{ {p}_{i} \}$?  Under these circumstances, marginal entropies do not increase on the average, i.e., $S({\rho}^{a}) \geq  {\sum}_{i}{p}_{i}S({\rho}^{a}_{i})$, where $\{ {\rho}^{a}_{i} \}$ are the marginal states of ${\psi}_{i}$ \cite{bpr}.  Not surprisingly, $C[\sigma(\psi)]$ turns out to be similarly monotonic.

To establish the above assertion, we first consider the non-negative quantity $f(\chi)={\sum}_{a=1}^{N} S({\rho}^{a})-{S}^{max}$, where $\chi$ is any pure composite, $\{ {\rho}^{a} \}$ its marginals, and ${S}^{max}$ the maximum of its marginal entropies.  This quantity can be reorganized as $f(\chi)={\min}_{\,b}[{\sum}_{a=1, a \neq b}^{N} S({\rho}^{a})]$, i.e., as the minimum of the sum of all but one marginal entropy.  Using the monotonicity condition for marginal entropies, we find $f(\chi) \geq {\min}_{\,b}[{\sum}_{a=1, a \neq b}^{N} {\sum}_{i}{p}_{i}S({\rho}^{a}_{i})]$.  The two sums in this equation can be reordered.  Moreover, interchanging the sum over $i$ with the minimum over $b$ in the resulting expression cannot make it any larger.  Effecting these changes, we find that $f(\chi) \geq {\sum}_{i}{p}_{i} f({\chi}_{i})$.  This result can be restated as

\textit{Theorem 3.  The sum of marginal entropies less their maximum is an LOCC monotone for any pure, multipartite state}.

Clearly, $f(\cdot)$ is a readily computable, pure state entanglement monotone, and as such provides a extension of the standard entropy of entanglement to multipartite states.  Returning to the multi-qubit state $\psi$, we observe that $C[\sigma(\psi)]=f(\psi)$, since the majorization and entropic orders are equivalent for qubits.  We can therefore use Theorem 3 to arrive at

\textit{Corollary 3. The sum of marginal entropies less the entropy of their infimum is an LOCC monotone for any multipartite pure state composed of qubits}.

It is a reasonable conjecture that Corollary 3 holds for marginals of arbitrary dimension, not just qubits.  If so, $C[\sigma(\cdot)]$ would represent an entanglement measure for all pure states.  Remarkably, the separable correlative capacity of the marginals of a pure state reliably track its entangled correlations.  Note, incidentally, that $f(\psi)$ provides an upper bound for $C[\sigma(\cdot)]$ since the majorization order implies the entropic one.

This work was supported in part by an award from California State University, Sacramento.
{}


\begin{thebibliography}{}
\bibitem{note} Note that the infimum or supremum of a set need not belong to the set.
\end{thebibliography}
\end{document}